\documentclass[twocolumn,prb,showpacs,preprintnumbers,amsmath,amssymb,floatfix]{revtex4}
\usepackage{graphicx}
\usepackage{amsmath}
\usepackage{amssymb}
\usepackage{color}

\begin{document}
%

\title{Weak and strong coupling limits of the two-dimensional Fr\"ohlich
polaron with spin-orbit Rashba interaction}

\author{C. Grimaldi}

\affiliation{Max-Planck-Institut f\"ur Physik komplexer Systeme,
N\"othnitzer Srt.38, D-01187 Dresden Germany \\
LPM, Ecole Polytechnique F\'ed\'erale de
Lausanne, Station 17, CH-1015 Lausanne, Switzerland}



\begin{abstract}
The continuous progress in fabricating low-dimensional systems with large
spin-orbit couplings has reached a point in which nowadays materials may display
spin-orbit splitting energies ranging from a few to hundreds of meV. This situation
calls for a better understanding of the interplay between the spin-orbit coupling
and other interactions ubiquitously present in solids, in particular when the spin-orbit
splitting is comparable in magnitude with characteristic energy scales such as the Fermi
energy and the phonon frequency.

In this article, the two-dimensional Fr\"ohlich electron-phonon problem is reformulated
by introducing the coupling to a spin-orbit Rashba potential, allowing for a description
of the spin-orbit effects on the electron-phonon interaction. The ground state of the
resulting Fr\"ohlich-Rashba polaron is studied in the weak and strong coupling limits
of the electron-phonon interaction for arbitrary values of the spin-orbit splitting.
The weak coupling case is studied within the Rayleigh-Schr\"odinger
perturbation theory, while the strong-coupling electron-phonon regime is investigated by
means of variational polaron wave functions in the adiabatic limit.
It is found that, for both weak and strong coupling polarons,
the ground state energy is systematically lowered by the spin-orbit interaction, indicating
that the polaronic character is strengthened by the Rashba coupling. It is also shown that,
consistently with the lowering of the ground state, the polaron effective mass is enhanced
compared to the zero spin-orbit limit. Finally, it is argued that the crossover between 
weakly and strongly coupled polarons can be shifted by the spin-orbit interaction.
\end{abstract}
\maketitle

\section{Introduction}
\label{intro}
The Fr\"ohlich Hamiltonian describing a single electron coupled to
longitudinal optical phonons is a paradigmatic model of the electron-phonon (el-ph)
interaction,\cite{fro} and has represented in the past, in addition to its interest for
the solid-state physics, an ideal problem
for testing many mathematical methods in quantum field theory.\cite{mitra}
Because of the coupling with the phonon field, the resulting quasi-particle,
the polaron, has an effective mass larger, and a ground state energy lower,
than the free electron. These quantities have been investigated for the
three-dimensional (3D) case by means of perturbation theory for the weak-coupling
limit,\cite{fpz} and of variational treatments for the intermediate,\cite{llp} and
strong-coupling cases.\cite{pekar,miyake} The path-integral variational calculations of
Feynman,\cite{feynman} and subsequent refinements of this method,\cite{ganbold}
have provided a solid description for all values of the
coupling, verified also by improved variational methods,\cite{defilippis} and
by quantum Monte-Carlo studies.\cite{alex,ciuck}

The interest aroused some time ago on semiconductor heterojunctions, or other low-dimensional
systems, prompted to modify the Fr\"ohlich model to accounting for two-dimensional
(2D) and quasi-2D systems.\cite{dassarma} By applying the same methods derived for the 3D case,
the ground state properties for the strictly 2D case were evaluated for weak, strong and intermediate
couplings,\cite{huy,devre1,devre2,qinghu} and the obtained systematic lowering of the ground state energy and
the enhancing of the effective mass compared to the 3D case has pointed out
the role of dimensionality in enhancing the polaronic character.\cite{dassarma,devrerev}

Concerning the el-ph problem in low dimensions,
recent progresses in developing high-quality low-dimensional systems
and in material engineering provide hints that, for a vast class of 
low-dimensional materials, the usual 2D Fr\"ohlich model, as considered 
in literature, may be incomplete. This concern comes
about by considering 2D systems exhibiting strong
spin-orbit (SO) splitting of the electronic states due to the inversion asymmetry
in the direction orthogonal to the conducting plane (Rashba SO mechanism).
This situation is encountered in semiconductor quantum wells with
asymmetric confining potentials,\cite{zutic} in the surface states of metals and
semi-metals,\cite{lashell,koroteev,sugawara}
and in surface alloys such as Li/W(110),\cite{rotenberg} Pb/Ag(111),\cite{pacile,ast1}
and Bi/Ag(111),\cite{astprl} with SO splitting energies ranging from a few meV in GaAs
quantum wells to about 0.2 eV in Bi/Ag(111).\cite{astprl}
In these systems, therefore, the SO energy may be of the same order or even much larger
than the typical phonon frequency, rising the question of how such state of affair affects
the el-ph interaction, in general, and the Fr\"ohlich coupling, in particular.

As pointed out in several
works,\cite{rashba,galstyan,grima,magarill,prl_super,cappe} the
Rashba interaction describing the SO coupling can have profound
effects on the low energy properties of the itinerant electrons.
Namely, in the low density regime, the Rashba SO coupling induces
a topological change of the Fermi surface of the free electrons,
leading to an effective reduction of the dimensionality in the
electronic density of states (DOS). In this situation, a 2D low
density electron gas would develop, in the presence of SO Rashba
coupling, a phenomenology similar to one-dimensional (1D) systems, 
triggered by the square-root divergence of the (effectively 1D)
DOS at low energies.

Some interesting consequences of this scenario on the el-ph
problem have already been discussed in
Ref.[\onlinecite{prl_super}], concerning the superconducting
transition, and in Ref.[\onlinecite{cappe}] for the effective
mass and the spectral properties. The picture arising from these
works, although being limited to the momentum-independent
Holstein el-ph interaction and to weak-to-moderate
couplings, confirms that, for sufficiently low electron densities,
the coupling to the phonons is amplified by the SO interaction
through the 1D-like divergence of the DOS.

Notwithstanding the relevance of these results for the Holstein
model, the use of a local el-ph interaction may however result
inadequate in the extremely low electron density regime,
where the SO effects are more evident.\cite{prl_super,cappe} Indeed,
the lack of effective screening in this case would rather suggest a long-range
interaction as being a more appropriate description of the el-ph
coupling. It is therefore natural to consider the 2D
Fr\"ohlich polaron, and its coupling to the SO interaction, as
a model better describing the unscreened el-ph interaction in 2D Rashba
systems in the low density limit.

In this article, a single electron moving with a parabolic
dispersion in the two-dimensional $x$-$y$ plane is coupled
simultaneously to the Rashba SO potential and to the phonon
degrees of freedom through a Fr\"ohlich interaction term. The
total system is then described by the 2D Fr\"ohlich-Rashba
Hamiltonian $H=H_{el}+H_{ph}+H_{el-ph}$, where ($\hbar=1$)
\begin{equation}
\label{hel}
H_{el}=\frac{p^2}{2m}+\mathbf{\Omega}(\mathbf{p})
\cdot\boldsymbol{\sigma}
\end{equation}
is the Hamiltonian for an electron with mass $m$ and momentum operator
${\bf p}=-i\boldsymbol{\nabla}$ with components $(p_x,p_y,0)$, $\mbox{\boldmath$\sigma$}$
is the spin-vector operator with components given by the Pauli matrices, and
$\mathbf{\Omega}(\mathbf{p})$ is the SO vector field which in the case of Rashba
coupling reduces to:
\begin{equation}
\label{hso}
\mathbf{\Omega}(\mathbf{p})=\gamma\left(
\begin{array}{c}
- p_y \\
 p_x \\
0 \end{array}\right),
\end{equation}
where $\gamma$ is the SO coupling parameter. The phonon part of the Hamiltonian
is given by
\begin{equation}
\label{hph}
H_{ph}=\omega_0\sum_{\bf q} a^\dagger_{\bf q}a_{\bf q},
\end{equation}
where $a^\dagger_{\bf q}$ ($a_{\bf q}$) is the creation
(annihilation) operator for a phonon with momentum ${\bf
q}=(q_x,q_y)$ and optical frequency $\omega_0$. The el-ph
interaction Hamiltonian for the 2D electron coupled to 
longitudinal optical (LO) phonons is:\cite{dassarma,devre1}
\begin{equation}
\label{helph}
H_{el-ph}=\frac{1}{\sqrt{A}}\sum_{\bf q}\frac{1}{\sqrt{q}} (M_0 e^{i{\bf q}\cdot{\bf r}}a_{\bf q}+
M_0^* e^{-i{\bf q}\cdot{\bf r}}a^\dagger_{\bf q})
\end{equation}
with
\begin{equation}
\label{Mq}
M_0=i\omega_0 \left(\frac{2\pi^2\alpha^2}{m\omega_0}\right)^{1/4},
\end{equation}
where
$\alpha=e^2(\epsilon_\infty^{-1}-\epsilon_0^{-1})\sqrt{m/2\omega_0}$
is the dimensionless el-ph coupling constant, with $e$ being the
electron charge, and $\epsilon_\infty$ and $\epsilon_0$ the high
frequency and static dielectric constants, respectively.

It is worth clarifying here the significance of the 2D Fr\"ohlich interaction of 
Eq.\eqref{helph} with respect to the characteristics of specific materials.
For quantum wells and 2D heterostructures, where the electron wave function
is assumed here to be confined in a sheet of zero thickness, Eq.\eqref{helph}
describes the coupling of the electron to bulk LO phonons, while the coupling to
interface phonon modes is neglected. The inclusion of such interface phonon 
contributions may be important in describing specific materials, but it is 
unnecessary for the present study, where the focus is on the SO effects on the unscreened
(long-range) el-ph interaction, for which Eq.\eqref{helph} is a paradigm for the 2D case.
Concerning the el-ph coupling of electronic surface states, Eq.\eqref{helph} coincides
(apart from a redefinition of $M_0$) with the coupling to 2D surface phonons when 
the coupling to bulk phonons extending 
below the surface is negligible.\cite{evans} Such approximation is coherent with the
ideal 2D assumption for the electron wave function, which is physically realized when
the electronic surface states have negligible coupling to the bulk.
A further motivation of using the 2D Fr\"ohlich model \eqref{helph} is that,
in the absence of SO interaction, the ground state polaron energy $E_P$ and
effective mass $m^*$ have already been studied by several 
authors,\cite{dassarma,huy,devre1,devre2,qinghu} and the exact results obtained for 
the weak ($\alpha\ll 1 $) and strong ($\alpha\gg 1$)
coupling limits provide a useful reference for the effect of nonzero SO coupling.

In the present work, the 2D Fr\"ohlich-Rashba Hamiltonian is studied by
considering the weak and strong coupling limits of the el-ph interaction,
with arbitrary strength of the SO coupling $\gamma$. For $\alpha\ll 1$
the polaron energy $E_P$ and the effective mass $m^*$ are obtained from 
second order perturbation theory in Sec.\ref{weak}, where numerical and exact 
analytical results are presented. It is
shown that the effect of $\gamma\neq 0$ is qualitatively similar
to that observed in the Holstein model,\cite{prl_super,cappe}
namely, the SO coupling enhances the effective coupling to the
phonons. In particular, $E_P$ is lowered by $\gamma$ and,
simultaneously, the effective mass $m^*$ is enhanced. 
In Sec.\ref{strong} the strong coupling limit $\alpha\gg 1$ is treated
by the variational method, providing a rigorous upper bound of the ground 
state energy for arbitrary values of the SO interaction. As for the weak el-ph
coupling case, it is found that $E_P$ ($m^*$) is lowered (enhanced) by the SO interaction,
implying that the Rashba coupling always amplifies the polaronic character, regardless
of whether the el-ph interaction is weak or strong.

\section{Weak coupling}
\label{weak} In the presence of SO interaction, the electron wave
function is a spinor and its Green's function is conveniently
represented by a $2\times 2$ matrix in the spin subspace. For
$\alpha=0$ the free electron Green's function ${\bf G}_0$ is
readily obtained from $H_{el}$:
\begin{eqnarray}
\label{green1}
{\bf G}_0({\bf k},\omega)&=&\left(\omega-\frac{k^2}{2m}-\mathbf{\Omega}(\mathbf{k})
\cdot\boldsymbol{\sigma}\right)^{-1} \nonumber \\
&=&\frac{1}{2}\sum_{s=\pm} [1+s\hat{\mathbf{\Omega}}(\mathbf{k})
\cdot\boldsymbol{\sigma}]G_0^s(k,\omega),
\end{eqnarray}
where ${\bf k}$ is a 2D electron wavenumber, $\hat{\mathbf{\Omega}}(\mathbf{k})=
\mathbf{\Omega}(\mathbf{k})/\vert\mathbf{\Omega}(\mathbf{k})\vert$ and
\begin{equation}
\label{green2}
G_0^s(k,\omega)=\frac{1}{\omega-k^2/2m -s\gamma k}
\end{equation}
is the free electron propagator for the two ($s=\pm 1$) chiral states
characterized by two distinct bands with shifted parabolic dispersions $k^2/2m\pm \gamma k$.
The lowest band has its minimum value $-E_0$ at $k=k_0$, where $k_0$ and
$E_0$ are the Rashba momentum and energy defined respectively by:
\begin{equation}
\label{k0E0}
k_0=m\gamma, \,\,\,\,\,\, E_0=\frac{m}{2}\gamma^2.
\end{equation}
For later convenience, it is useful to express the electron
energy relative to $E_0$, so that the poles of Eq.(\ref{green2})
appear at energies:
\begin{equation}
\label{green3}
E_\pm(k)=\frac{1}{2m}(k\pm k_0)^2.
\end{equation}
The free electron ground state
is then given by the electron occupying the lower band at wavenumber $k=k_0$
with energy $\omega=0$.

In the weak el-ph coupling limit ($\alpha\ll 1$) the ground state
properties are obtained by the electron self-energy evaluated in
the second order perturbation theory. At zero temperature, the
resulting single electron self-energy is therefore:
\begin{equation}
\label{self1}
\mathbf{\Sigma}({\bf k},\omega)=\vert M_0\vert^2\int\!\frac{d{\bf k}'}{(2\pi)^2}
\frac{1}{\vert{\bf k}-{\bf k}'\vert}{\bf G}_0({\bf k}',\omega-\omega_0).
\end{equation}
Because of the momentum dependence of the Fr\"ohlich interaction, and
contrary to the Holstein el-ph case considered in Ref.[\onlinecite{cappe}],
the self-energy is not diagonal in the spin subspace. However, since the
momentum dependence enter only through the modulus of the momentum transfer,
equation (\ref{self1}) can be rewritten in a quite simple form. By using
$(\hat{\mathbf{\Omega}}(\mathbf{k})\cdot\boldsymbol{\sigma})^2=1$ and
\begin{equation}
\label{trick}
(\hat{\mathbf{\Omega}}(\mathbf{k})\cdot\boldsymbol{\sigma})
(\hat{\mathbf{\Omega}}(\mathbf{k}')\cdot\boldsymbol{\sigma})
=\hat{{\bf k}}\cdot\hat{{\bf k}}'+
(\hat{{\bf k}}\times\hat{{\bf k}}')\sigma_x\sigma_y,
\end{equation}
then the quantity $\hat{\mathbf{\Omega}}(\mathbf{k}')\cdot\boldsymbol{\sigma}$
appearing in Eq.(\ref{self1}) through ${\bf G}_0({\bf k}',\omega-\omega_0)$
can be replaced simply by $(\hat{\mathbf{\Omega}}(\mathbf{k})\cdot\boldsymbol{\sigma})
\,\hat{{\bf k}}\cdot\hat{{\bf k}}'$ because the second term of Eq.(\ref{trick})
vanishes after the integration over ${\bf k}'$. In this way, the resulting self-energy
reduces to:
\begin{equation}
\label{self2}
\mathbf{\Sigma}({\bf k},\omega)=\Sigma_d(k,\omega)\mathbf{1}+
\Sigma_o(k,\omega)\hat{\mathbf{\Omega}}(\mathbf{k})\cdot\boldsymbol{\sigma},
\end{equation}
where $\mathbf{1}$ is the unit matrix and $\Sigma_d$ and $\Sigma_o$ are,
respectively, the diagonal and off-diagonal contributions to the self-energy, both
depending solely on the modulus of ${\bf k}$.\cite{noteself} Their explicit expressions are:
\begin{align}
\label{self3a}
\Sigma_d(k,\omega)&\!=\!\frac{\vert M_0\vert^2}{2}\!\int\!\!\frac{d{\bf k}'}{(2\pi)^2}
\sum_s\frac{1}{\vert{\bf k}-{\bf k}'\vert}\frac{1}{\omega-\omega_0-E_s(k')},\\
\label{self3b}
\Sigma_o(k,\omega)&\!=\!\frac{\vert M_0\vert^2}{2}\!\int\!\!\frac{d{\bf k}'}{(2\pi)^2}
\sum_s\frac{1}{\vert{\bf k}-{\bf k}'\vert}
\frac{s{\bf k}\cdot{\bf k}'}{\omega-\omega_0-E_s(k')}.
\end{align}
In the limit of zero SO coupling, since $E_s(k)\rightarrow k^2/2m$,
$\Sigma_o(k,\omega)$ vanishes because of the summation over
$s=\pm 1$ in Eq.\eqref{self3b}. Notice also that, independently of $\gamma$,
$\Sigma_o(k,\omega)=0$ when the factor $1/\vert{\bf k}-{\bf k}'\vert$ in
Eq.\eqref{self3b} is replaced by a constant, as in the momentum-independent
Holstein el-ph coupling model.

By using Eq.(\ref{self2}) the Dyson equation for the interacting propagator ${\bf G}$
reduces to
\begin{eqnarray}
\label{self4}
{\bf G}^{-1}({\bf k},\omega)&=&{\bf G}_0^{-1}({\bf k},\omega)-\mathbf{\Sigma}({\bf k},\omega)
\nonumber \\
&=&\omega-\frac{k^2}{2m}-\Sigma_d(k,\omega)-E_0\nonumber \\
&&-[\gamma k +\Sigma_o(k,\omega)]
\hat{\mathbf{\Omega}}(\mathbf{k})\cdot\boldsymbol{\sigma},
\end{eqnarray}
and the poles $\omega_\pm$ of ${\bf G}$ are then given by:
\begin{equation}
\label{poles}
\omega_\pm=E_\pm(k)+\Sigma_d(k,\omega_\pm)\pm\Sigma_o(k,\omega_\pm).
\end{equation}
Now, the Rayleigh-Schr\"odinger perturbation theory permits to evaluate
the lower energy pole $\omega_-$ at the lowest order in the el-ph coupling $\alpha$.
This is accomplished by replacing $\omega_-$ by the unperturbed energy $E_-(k)$ in
the energy variables of $\Sigma_d$ and $\Sigma_o$. In this way, the lower pole
reduces to $\omega_-=E_-(k)+\Sigma_-(k)+\mathcal{O}(\alpha^2)$, where
\begin{equation}
\label{self5}
\Sigma_-(k)=\Sigma_d(k,E_-(k))-\Sigma_o(k,E_-(k)).
\end{equation}
Finally, by expanding $\Sigma_-(k)$ up to the second order in $k-k_0$,
the polaron dispersion in the vicinity of $k_0$ can be written as:
\begin{equation}
\label{w1}
\omega_-=E_P+\frac{1}{2m^*}(k-k_0^*)^2,
\end{equation}
where the polaron ground-state energy $E_P$, the effective mass $m^*$, and
the effective Rashba momentum $k_0^*$ are given respectively by:
\begin{align}
\label{EP1}
E_P&=\Sigma_-(k_0)-\frac{m^*}{2}\Sigma_-'(k_0)^2\nonumber \\
&=\Sigma_-(k_0)+\mathcal{O}(\alpha^2),\\
\label{m1}
\frac{m^*}{m}&= [1+m\Sigma_-''(k_0)]^{-1}\nonumber \\
&=1-m\Sigma_-''(k_0)+\mathcal{O}(\alpha^2),\\
\label{k01}
\frac{k_0^*}{k_0}&= 1-\frac{m^*}{k_0}\Sigma_-'(k_0)\nonumber \\
&= 1-\frac{m}{k_0}\Sigma_-'(k_0)+\mathcal{O}(\alpha^2).
\end{align}

\begin{figure}[b]
\protect
\includegraphics[scale=0.7,clip=true]{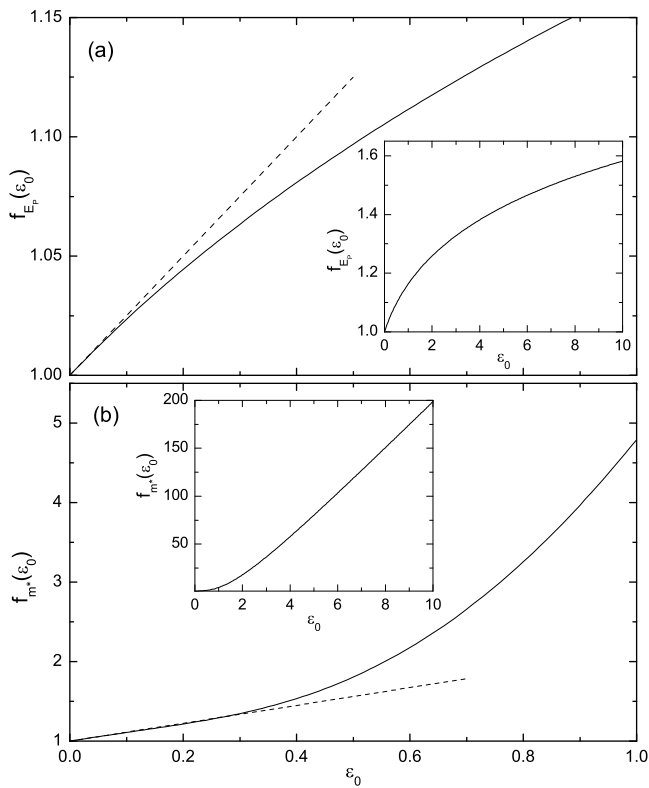}
\caption{(a): ground state energy factor $f_{E_P}(\varepsilon)$
as a function of the SO parameter $\varepsilon_0=E_0/\omega_0$.
The solid line is the numerical calculation, while the dashed
line is the weak SO limit Eq.\eqref{EP3}. Inset:
$f_{E_P}(\varepsilon)$ plotted for a wider range of
$\varepsilon_0$. (b): the effective mass factor
$f_{m^*}(\varepsilon_0)$ from numerical integration (solid line)
and from Eq.\eqref{m3} (dashed line). Inset:
$f_{m^*}(\varepsilon_0)$ plotted for a wider range of
$\varepsilon_0$. } \label{fig1}
\end{figure}

Let us first consider $E_P$ and $m^*$. In the zero SO limit, Eqs.
\eqref{EP1} and \eqref{m1} at $k_0=0$ lead respectively to
$E_P=\pi\alpha\omega_0/2$ and $m^*/m=1+\pi\alpha/8$, which
correspond to the results already reported in
Refs.[\onlinecite{huy,devre1,devre2}]. For finite values of the SO
coupling the ground state energy and the effective mass can be
expressed as
\begin{align}
\label{EP2}
E_P&=-\frac{\pi}{2}\alpha\omega_0 f_{E_P}(\varepsilon_0),\\
\label{m2} \frac{m^*}{m}&=1+\frac{\pi}{8}\alpha
f_{m^*}(\varepsilon_0),
\end{align}
where the factors $f_{E_P}(\varepsilon_0)$ and
$f_{m^*}(\varepsilon_0)$ contain all the effects of the SO
interaction and depend solely on the dimensionless SO parameter
\begin{equation}
\label{ve0}
\varepsilon_0\equiv\frac{E_0}{\omega_0}=\frac{m\gamma^2}{2\omega_0}.
\end{equation}
In the weak SO limit, the self-energy terms \eqref{self3a} and
\eqref{self3b} can be expanded in powers of the SO interaction,
allowing for an analytical evaluation of the integrals.
In this way, up to the linear order in $\varepsilon_0$,
$f_{E_P}(\varepsilon_0)$ and $f_{m^*}(\varepsilon_0)$ are found to be:
\begin{align}
\label{EP3}
f_{E_P}(\varepsilon_0)&=1+\frac{\varepsilon_0}{4}+\mathcal{O}(\varepsilon_0^2),\\
\label{m3}
f_{m^*}(\varepsilon_0)&=1+\frac{9}{8}\varepsilon_0+\mathcal{O}(\varepsilon_0^2),
\end{align}
indicating that the polaronic character is strengthened by the SO interaction
since, through Eqs. \eqref{EP2} and \eqref{m2}, the polaron energy $E_P$
is lowered and, simultaneously, the effective mass $m^*$ is enhanced
when $\varepsilon_0>0$.
This feature is not limited to the small $\varepsilon_0$ limit, but
holds true for arbitrary strengths of the SO coupling. This is
shown in Fig. \ref{fig1} where $f_{E_P}(\varepsilon_0)$ and
$f_{m^*}(\varepsilon_0)$, obtained from a numerical integration
of Eqs.\eqref{self3a} and \eqref{self3b}, are plotted as a
function of $\varepsilon_0$ by solid lines and compared with Eqs.
\eqref{EP3} and \eqref{m3} (dashed lines). The same quantities
calculated for a wider range of $\varepsilon_0$ are plotted in
the insets of Fig.\ref{fig1} and confirm that the ground state
energy $E_P$ and the effective mass $m^*$ are continuous functions
of $\varepsilon_0$ and are, respectively, further lowered and
enhanced by the SO coupling. In the strong SO limit
$\varepsilon_0\gg 1$, it is found that $f_{E_P}(\varepsilon_0)$
grows as $\ln(\varepsilon_0)$  while $f_{m^*}(\varepsilon_0)$
grows linearly. It is interesting to note that the Holstein-Rashba
model studied in Ref.[\onlinecite{cappe}] predicts results qualitatively
similar to the Fr\"ohlich model, indicating that the SO interaction
strengthen the polaronic character
independently of the specific form of the el-ph interaction.\cite{massholstein}

\begin{figure}[t]
\protect
\includegraphics[scale=0.69,clip=true]{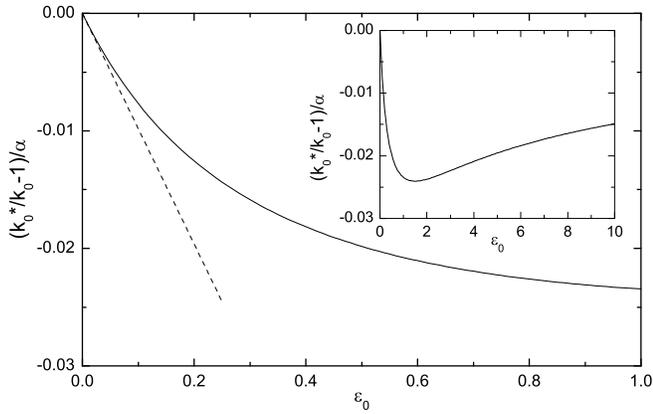}
\caption{Effective Rashba momentum $k_0^*$ as a function of the
SO parameter $\varepsilon_0=E_0/\omega_0$. The numerical
integration of Eq.\eqref{k01} (solid line) is compared with the
weak SO result \eqref{k02} (dashed line). Inset: the same
quantity plotted for a wider range of $\varepsilon_0$.}
\label{fig2}
\end{figure}

In addition to $E_P$ and $m^*$, the interplay between the el-ph coupling and the
SO interaction modifies also the Rashba momentum $k_0$ through Eq.\eqref{k01}.
In the weak SO limit, the effective quantity $k_0^*$ is found to be
\begin{equation}
\label{k02}
\frac{k_0^*}{k_0}\simeq 1-\frac{\pi}{32}\alpha\varepsilon_0,
\end{equation}
indicating a reduction of the bare Rashba momentum $k_0$, confirmed also
by the numerical calculation of Eq.\eqref{k01} reported in Fig. \ref{fig2} by the
solid line. As shown in the inset, for fixed el-ph coupling $\alpha$, $k_0^*$ however
does not deviate much from its bare limit $k_0$, even for large values of the
SO parameter $\varepsilon_0$.

Let us compare now the present results with those appeared recently in
literature. In Ref.[\onlinecite{liu}] the ground state energy of a polaron 
near a polar-polar semiconductor interface with Rashba SO coupling has
been evaluated with the Lee-Low-Pines method.\cite{llp} As a function of the
SO splitting, the polaron ground state is found to be lowered, in qualitative
agreement therefore with the present results. A more quantitative comparison 
is however precluded by the different model of Ref.[\onlinecite{liu}], where
contributions from interface phonon modes and confining potentials are
considered as well. In another work,\cite{li} the Rayleigh-Schr\"odinger 
perturbation theory has been applied to the polaron ground state of the 
2D Fr\"ohlich-Rashba model, permitting therefore a direct comparison with 
the analysis presented here. Despite that the authors of Ref.[\onlinecite{li}]
find that the polaron ground state is lowered by $\varepsilon_0$, their
values of $E_P$ differ from those plotted in Fig. \ref{fig1}(a). 
In Ref.[\onlinecite{li}], in fact, the ground state energy factor $f_{E_P}$ is
found to be $f_{E_P}(\varepsilon_0)=1/\sqrt{1-\varepsilon_0}$, which implies 
a small $\varepsilon_0$ expansion different from Eq.\eqref{EP3} and, more 
importantly, a divergence of $E_P$ at $\varepsilon_0=1$. 
In Fig. \ref{fig1}(a), instead, nothing of special happens at $\varepsilon_0=1$.
This discrepancy is easily traced back in the fact that in Ref.[\onlinecite{li}]
the expansion of $\Sigma_-(k)$, Eq.\eqref{self5}, is made around $k=0$, instead of $k=k_0$
as done here, which does not correspond to a perturbative calculation 
of the ground-state energy.

The results presented in this section have been derived by assuming a weak coupling
to the phonons. However, as it is clear from the plots in Fig. \ref{fig1}, the enhancement
of the polaronic character driven by $\varepsilon_0$ for fixed $\alpha$ unavoidably renders
the perturbative approach invalid for sufficiently large $\varepsilon_0$ values. For example,
from Eq.\eqref{m2}, the validity of the weak coupling results for $m^*/m$ are subjected to
the condition $\alpha f_{m^*}(\varepsilon_0)\ll 1$, otherwise higher order el-ph contributions
should be considered for a consistent description of the SO effects.
The question remains therefore whether the SO enhancement of the polaronic character survives
also for large $\alpha$ values, or it is instead limited to the weak coupling limit. In the
next section, this problem is studied for the limiting case of strong el-ph interaction
$\alpha\gg 1$, providing therefore, together with the weak coupling results, a global
understanding of the SO effects on the Fr\"ohlich polaron.

\section{Strong coupling}
\label{strong}

It is well known that a perturbative scheme such that employed in the previous section
fails to describe the Fr\"ohlich polaron ground state when the el-ph coupling is very large.
This is due to the fact that for $\alpha \gg 1$ the lattice polarization, and resulting
``self-trapping" effect experienced by the the electron,\cite{note_trap} renders the plane
wave representation of
the unperturbed electron inappropriate for obtaining the polaron ground state.
Instead, as originally proposed in Ref.[\onlinecite{pekar}] and rigorously proved in
Refs.[\onlinecite{donsker,lieb}], the asymptotic description of
the polaron wave function in the strong
coupling limit $\alpha\gg 1$ is that of a product between purely electronic, $\psi({\bf r})$,
and purely phononic, $\vert\xi\rangle$, wave functions.
Within such adiabatic limit, the ground state energy and the effective mass of a
2D Fr\"ohlich polaron have been calculated in Refs.[\onlinecite{devre1,devre2}] by using
the variational method with different ansatz wave functions. From Ref.[\onlinecite{devre1}],
one realizes that
exponential, gaussian and Pekar-type wave functions provide increasingly better estimates of $E_P$
with accuracies respectively of $14\%$,  $0.3\%$, and $0.03\%$ with respect to the exact
ground state energy $E_P/\omega_0=-0.40474\alpha^2$, obtained by a numerical solution of
the integro-differential equation for the electron wave function.\cite{qinghu}
In the following, the variational method is used to evaluate the SO effects
on the polaron ground state.

\subsection{trial wave functions}
\label{wave}
For the nonzero SO case, due to the presence of the Pauli matrices in  Eq.\eqref{hel},
suitable ansatz wave functions
must take into account the electron spin degrees of freedom. Hence, in full generality,
the strong-coupling polaron wave function may be represented as:
$\vert\mathbf{\Psi},\xi\rangle=\mathbf{\Psi}({\bf r})\vert \xi\rangle$,
where $\mathbf{\Psi}({\bf r})$ is a two-components spinor for the electron.
The corresponding expectation value of the total Hamiltonian $H$ is:
\begin{align}
\label{sc1}
\langle\mathbf{\Psi},\xi\vert H\vert\mathbf{\Psi},\xi\rangle =&
\langle\mathbf{\Psi}\vert H_{el}\vert\mathbf{\Psi}\rangle+
\langle\xi\vert H_{ph}\vert\xi\rangle \nonumber \\
&+\frac{1}{\sqrt{A}}\sum_{\bf q}\frac{1}{\sqrt{q}}(M_0\rho({\bf q})
\langle\xi\vert a_{\bf q}\vert\xi\rangle+{\rm h.c.}),
\end{align}
where
\begin{equation}
\label{rho1}
\rho({\bf q})=\langle\mathbf{\Psi}\vert e^{i{\bf q}\cdot{\bf r}}\vert
\mathbf{\Psi}\rangle
=\int\!d{\bf r} e^{i{\bf q}\cdot{\bf r}} \vert\mathbf{\Psi}({\bf r})\vert^2.
\end{equation}
The form of Eq.\eqref{sc1} permits to integrate out the phonon wave function in the usual way.
Hence, by introducing the phonon coherent state $\vert\xi\rangle=
\mathcal{N}e^{\sum_{\bf q}\xi_{\bf q} a^\dagger_{\bf q}}\vert0\rangle$, where $\mathcal{N}$
is a normalization factor and $\xi_{\bf q}$ a variational parameter, minimization
of \eqref{sc1} with respect to $\xi_{\bf q}$ leads to the functional
\begin{equation}
\label{sc2}
E[\mathbf{\Psi}]=\langle\mathbf{\Psi}\vert H_{el}\vert\mathbf{\Psi}\rangle
-\frac{\vert M_0\vert^2}{\omega_0}\int\!\frac{d{\bf q}}{(2\pi)^2}\frac{1}{q}\vert\rho({\bf q})\vert^2,
\end{equation}
where the continuum limit $A^{-1}\sum_{\bf q}\rightarrow\int\!d{\bf q}/(2\pi)^2$ has
been performed. By choosing an appropriate functional form for $\mathbf{\Psi}({\bf r})$, and by
minimizing $E[\mathbf{\Psi}]$ with respect to the variational parameters defining $\mathbf{\Psi}({\bf r})$,
an upper bound for the ground state energy is then $E[\mathbf{\Psi}_0]$, where $\mathbf{\Psi}_0({\bf r})$
is such that $E[\mathbf{\Psi}_0]={\rm min}(E[\mathbf{\Psi}])$. 
As done in the previous section, the polaron energy is then obtained from
\begin{equation}
\label{EPans}
E_P=E[\mathbf{\Psi}_0]+E_0,
\end{equation}
where $E_0$ is the free-electron SO energy defined in Eq.\eqref{k0E0}.

Of course, the functional
form of $\mathbf{\Psi}({\bf r})$ is decisive for obtaining accurate
estimates of the ground state energy, and a suitable choice must be guided by looking at the
properties of the true ground state spinor $\mathbf{\Psi}_G({\bf r})$. These can be deduced
by a formal minimization of the functional $E[\mathbf{\Psi}]$ with respect to $\mathbf{\Psi}$.
By introducing the Lagrange multiplier $\epsilon$ to ensure
that the wave function is normalized to unity, minimization of \eqref{sc2} leads to:
\begin{equation}
\label{A1}
H_{el}\mathbf{\Psi}({\bf r})+V({\bf r})\mathbf{\Psi}({\bf r})=\epsilon\mathbf{\Psi}({\bf r}),
\end{equation}
where, by using the definition of $\rho({\bf q})$ given in in Eq.\eqref{rho1}:
\begin{align}
\label{A2}
V({\bf r})&=-\frac{2\vert M_0\vert^2}{\omega_0}\int\!\frac{d{\bf q}}{(2\pi)^2}
\frac{\rho({\bf q})^*}{q}e^{i{\bf q}\cdot{\bf r}}\nonumber \\
&=-\frac{\vert M_0\vert^2}{\pi\omega_0}\int\!d{\bf r}'
\frac{\vert \mathbf{\Psi}({\bf r}')\vert^2}{\vert {\bf r}-{\bf r}'\vert}.
\end{align}
From the above expression of $V({\bf r})$, the functional \eqref{sc2} can be
rewritten as $E[\mathbf{\Psi}]=\langle\mathbf{\Psi}\vert H_{el}\vert\mathbf{\Psi}\rangle
+\bar{V}/2$, where $\bar{V}=\langle\mathbf{\Psi}\vert V({\bf r})\vert\mathbf{\Psi}\rangle$.
Now, if $\mathbf{\Psi}_G$ is the exact
ground state wave function, with ground state energy $E_G=E[\mathbf{\Psi}_G]$, then, from \eqref{A1}
and $E_G=\langle\mathbf{\Psi}\vert H_{el}\vert\mathbf{\Psi}\rangle
+\bar{V}/2$, it is found that $\epsilon=E_G+\bar{V}/2$, so that Eq.\eqref{A1} reduces to:
\begin{equation}
\label{A3}
H_{el}\mathbf{\Psi}_G({\bf r})+[V({\bf r})-\bar{V}/2]\mathbf{\Psi}_G({\bf r})=E_G\mathbf{\Psi}_G({\bf r}).
\end{equation}
As noted in Ref.[\onlinecite{magarill}] (see also Refs.[\onlinecite{bulgakov,tsitsi}]),
the ground-state wave function of a 2D
electron subjected to a SO Rashba interaction and to a 2D central potential
({\it i.e.} a potential depending only upon $r=\vert {\bf r}\vert$) is of the form
\begin{equation}
\label{A4}
\mathbf{\Psi}_G({\bf r})=\left(
\begin{array}{l}
\psi_1(r) \\
\psi_2(r)\,e^{i\varphi}
\end{array}\right),
\end{equation}
where $\varphi$ is the azimuthal angle of ${\bf r}$. Now, if Eq.\eqref{A4} is used in Eq.\eqref{A2},
the resulting self-consistent potential depends only upon $r$, $V({\bf r})\rightarrow V(r)$,
so that Eq.\eqref{A4} is consistently also the correct form for the polaron ground-state wave function.
Hence, passing to polar coordinates, Eq.\eqref{A3} can be rewritten as a system of integro-differential
equations for the spinor components $\psi_1$ and $\psi_2$:
\begin{widetext}
\begin{align}
\label{dif1}
&\left[-\frac{1}{2m}\left(\frac{d^2}{d r^2}+\frac{1}{r}\frac{d}{dr}\right)+U(r)\right]\psi_1(r)
-\gamma\left(\frac{d}{dr}+\frac{1}{r}\right)\psi_2(r)=E_G\psi_1(r), \\
\label{dif2}
&\left[-\frac{1}{2m}\left(\frac{d^2}{d r^2}+\frac{1}{r}\frac{d}{dr}-\frac{1}{r^2}\right)
+U(r)\right]\psi_2(r)
+\gamma\frac{d}{dr}\psi_1(r)=E_G\psi_2(r),
\end{align}
\end{widetext}
where $U(r)=V(r)-\bar{V}/2$ and the polaron energy is obtained from $E_P=E_G+E_0$.
By introducing the dimensionless variable $\rho=r/\ell_P$, where $\ell_P=1/\alpha (m\omega_0)^{1/2}$
is a measure of the polaron spatial extension in the zero SO limit, and by noticing that $E_G$ does
not depend on the sign of $\gamma$, it is straightforward to realize from Eqs.\eqref{dif1} and
\eqref{dif2} that the polaron ground state energy scales as
\begin{equation}
\label{formEp}
E_P=\mathcal{F}\!\left(\frac{\varepsilon_0}{\alpha^2}\right) \alpha^2\omega_0,
\end{equation}
where $\varepsilon_0=E_0/\omega_0$ is the
dimensionless SO energy introduced in Eq.\eqref{ve0} and $\mathcal{F}$ is a generic function.
It is found therefore from Eq.\eqref{formEp} that the dependence of $E_P$ on the SO interaction 
is through the effective parameter $\varepsilon_0/\alpha^2$, which is treated in the following 
as an independent variable. Although $\varepsilon_0/\alpha^2$
is then formally allowed to vary from $0$ to $\infty$,
it is nevertheless important to estimate the range
over which $\varepsilon_0/\alpha^2$ is expected to vary for reasonable values of the microscopic
parameters $E_0$, $\omega_0$, and $\alpha$. To this end, it must be reminded that the strong
coupling limit of a 2D Fr\"ohlich polaron (in the absence of SO interaction) is
appropriate only for $\alpha\gtrapprox 5$,\cite{dassarma} and that the typical
phonon energy scale is of the order of few to tens meV, say $\omega_0\approx 5-10$ meV.
The largest value of the Rashba energy $E_0$ reported so far is of about $0.2$ eV,\cite{astprl}
so that $\varepsilon_0/\alpha^2\lesssim 1-2$ is a rather conservative estimate compatible
with material parameters and with the strong coupling polaron hypothesis.

Let us now evaluate the behavior of $\psi_1(r)$ and $\psi_2(r)$ for $r\ll \ell_P$ and
$r\gg \ell_P$. By requiring a regular solution at the origin, it turns out by 
inspection of Eqs.\eqref{dif1} and \eqref{dif2}
that the spinor components of \eqref{A4} behave as $\psi_1(r)={\rm const.}$ and
$\psi_1(r)\propto r$ as $r\rightarrow 0$, while the behavior for $r\gg \ell_P$ is
obtained from the large $r$ limit of Eqs.\eqref{dif1} and \eqref{dif2}:
\begin{align}
\label{dif3}
&-\frac{1}{2m}\frac{d^2\psi_1(r)}{d r^2}
-\gamma\frac{d\psi_2(r)}{dr}=W\psi_1(r), \\
\label{dif4}
&-\frac{1}{2m}\frac{d^2\psi_2(r)}{d r^2}+\gamma\frac{d\psi_1(r)}{dr}=W\psi_2(r),
\end{align}
where the quantity $W=E_G+\bar{V}/2$ is negative for bound states. Solutions
of Eqs.\eqref{dif3} and \eqref{dif4} which are finite for $r\rightarrow \infty$
are linear combination of $\exp(-\lambda_+r)$ and $\exp(-\lambda_-r)$ with
\begin{equation}
\label{lambda}
\lambda_\pm=\sqrt{-2m(E_P+\bar{V}/2)}\pm i k_0,
\end{equation}
implying an exponential decay of the polaron wave function, accompanied
by periodic oscillations of wavelength $2\pi/k_0$.

The informations gathered on the limiting behaviors of the ground state
wave function are sufficient for guessing some appropriate trial wave functions to be used
in Eq.\eqref{sc2}. By assuming that for zero SO coupling the electron is in a
spin-up state, then a simple ansatz compatible with the limits discussed above is
\begin{equation}
\label{ansatz}
\mathbf{\Psi}({\bf r})=f(r)\left(
\begin{array}{l}
\cos(b r)\\
\sin(b r)\,e^{i\varphi}
\end{array}\right),
\end{equation}
where $b$ is a variational SO parameter vanishing for $\gamma=0$ and
$f(r)$ is an exponentially decaying function for $r\rightarrow\infty$ and
such that $f(0)\neq 0$. The advantage of Eq.\eqref{ansatz} is that one
can use exponential or Pekar-type functions for $f(r)$, automatically recovering
therefore the known results for the zero SO case.\cite{devre1}  It should be noted, however, that in the
$U(r)\rightarrow 0$ limit Eq.\eqref{ansatz} does not reproduce correctly the behavior
of the exact ground state wave function, which is instead given by Eq.\eqref{A4}
with $\psi_1(r)$ and $\psi_2(r)$ proportional to the Bessel functions $J_0(k_0r)$ and
$J_1(k_0r)$, respectively.\cite{bulgakov,tsitsi} Hence, Eq.\eqref{ansatz} is not expected to provide
a reliable ground state energy in the strong SO regime, for which $U(r)$ can be
treated as a perturbation. To remedy to this deficiency, the following alternative
form of the polaron ansatz is proposed:
\begin{equation}
\label{ansatzB}
\mathbf{\Psi}({\bf r})=f(r)\left(
\begin{array}{l}
J_0(b r)\\
J_1(b r)\,e^{i\varphi}
\end{array}\right),
\end{equation}
where, as before, $b$ is a variational SO parameter.
As it will be shown below, the lowest value of $E_P$ is
given either by Eq.\eqref{ansatz} or by Eq.\eqref{ansatzB}, depending on the
specific form considered for $f(r)$ and on the value of the SO coupling.

\subsection{ground state energy}
\label{ground}

\begin{figure*}[t]
\protect
\includegraphics[scale=0.85,clip=true]{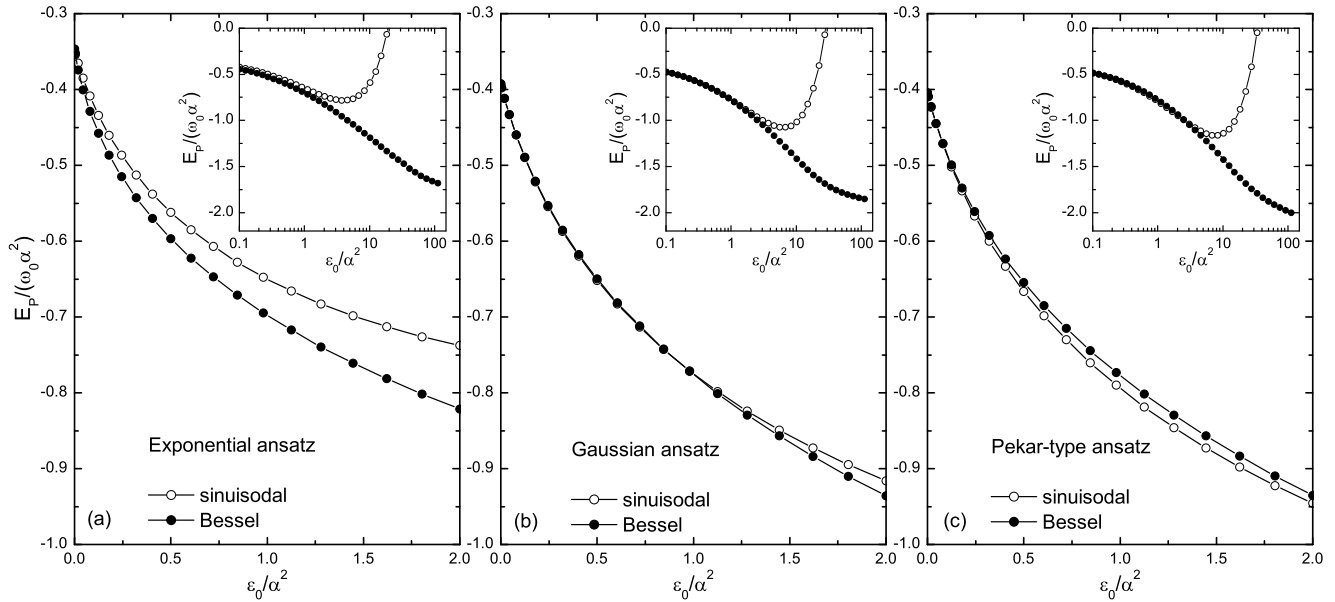}
\caption{Polaron ground state energy as a function
of $\varepsilon_0/\alpha^2$ for different trial wave functions for $f(r)$.
(a): exponential; (b): Gaussian; (c): Pekar.
The sinuisodal and the Bessel type of anstazes are given respectively by
Eq.\eqref{ansatz} and Eq.\eqref{ansatzB}. Inset: the polaron energy
for a wider range of $\varepsilon_0/\alpha^2$ values.} \label{fig3}
\end{figure*}

To evaluate the polaron ground state energy, three different trial wave functions
for $f(r)$ are considered: exponential, Gaussian and Pekar-type. As shown below,
the Gaussian ansatz will provide results comparable to those coming from the exponential
and Pekar functions, despite its faster decay for $r\rightarrow \infty$ compared
to Eq.\eqref{lambda}.
These three trial wave functions will be used
in combination with the sinuisodal and the Bessel-type spinors
of Eqs.\eqref{ansatz} and \eqref{ansatzB}, respectively, giving a total of six
different ansatzes for the Fr\"ohlich-Rashba polaron wave function.

\textit{Exponential ansatz}. Let us start by evaluating the functional
$E[\mathbf{\Psi}]$, Eq.\eqref{sc2}, by using the exponential
ansatz $f(r)=\mathcal{A}\exp(-ar)$, where $a$ is a variational parameter
and $\mathcal{A}$ is a normalization factor, in combination with the
sinuisodal trial wave function \eqref{ansatz}.
By introducing the dimensionless quantities $\tilde{a}=a \ell_P$,
$\tilde{b}=b \ell_P$, and $\tilde{\gamma}=k_0\ell_P$, for nonzero
SO interaction the functional \eqref{sc2} evaluated with the 
exponential-sinuisodal ansatz reduces to
\begin{align}
\label{expo1}
\frac{E[\mathbf{\Psi}]}{\alpha^2\omega_0}=&\frac{1}{2}
\left[\tilde{a}^2+\tilde{b}^2+\tilde{a}^2
\ln\!\left(1+\frac{\tilde{b}^2}{\tilde{a}^2}\right)\right]
\nonumber \\
&-\tilde{\gamma}\tilde{b}\left(1+\frac{\tilde{a}^2}
{\tilde{a}^2+\tilde{b}^2}\right)-\frac{3\pi \tilde{a}}{8\sqrt{2}}.
\end{align}
For weak SO couplings, Eq.\eqref{expo1} has its minimum at
$\tilde{b}=\tilde{\gamma}=\sqrt{2\varepsilon_0}/\alpha$ and
$\tilde{a}=3\sqrt{2}\pi/16$, so that the resulting polaron energy
$E_P=E[\mathbf{\Psi}_0]+E_0$ becomes
\begin{equation}
\label{expo2}
\frac{E_P}{\alpha^2\omega_0}=
-\left(\frac{3\pi}{16}\right)^2-\frac{\varepsilon_0}{\alpha^2}
+\mathcal{O}\!\left(\frac{\varepsilon_0^2}{\alpha^4}\right).
\end{equation}
In the $\varepsilon_0=0$ limit, Eq.\eqref{expo2} reduces to
$E_P/\alpha^2\omega_0=-(3\pi/16)^2\simeq-0.3469$, recovering
therefore the result of Ref.[\onlinecite{devre1}], while for
$\varepsilon_0>0$ the polaron energy is lowered by the SO
interaction, in qualitative analogy with the weak electron-phonon
behavior discussed in Sec.\ref{weak}. The lowering of $E_P$ is
confirmed by a numerical minimization of Eq.\eqref{expo1} whose
results are plotted in Fig.\ref{fig3}(a) (open circles). For
$\varepsilon_0/\alpha^2=1$, the polaron energy has dropped to
$E_P/\alpha^2\omega_0\simeq -0.65$, that is about two times lower
than the zero SO case. However, upon increasing
$\varepsilon_0/\alpha^2$, $E_P$ displays a minimum at
$\varepsilon_0/\alpha^2\simeq 3.98$ [inset of Fig.\ref{fig3}(a)]
and for larger values of the SO interaction the polaron energy
increases. Eventually, for $\varepsilon_0/\alpha^2\gtrsim 14$ the
calculated ground state energy becomes larger than the zero SO
value $E_P/\alpha^2\omega_0=-(3\pi/16)^2$. Such upturn of $E_P$
for large $\varepsilon_0$ stems from the inadequacy of the
sinuisodal components of \eqref{ansatz} in treating the
oscillatory behavior in the strong SO regime which, as pointed
out above, should instead be given by Bessel-type functions.
Indeed when the exponential ansatz for $f(r)$ is used in
Eq.\eqref{ansatzB}, rather than in Eq.\eqref{ansatz}, not only
the resulting $E_P$ is lower than the previous case, but also the
upturn of $E_P$ disappears, leading to a monotonous
lowering of the polaron energy as $\varepsilon_0/\alpha^2$
increases [filled circles in Fig.\ref{fig3}(a)].
As $\varepsilon_0/\alpha^2\rightarrow\infty$, however, the polaron energy
does not decrease indefinitely but rather approaches a limiting value.
Although an accurate numerical evaluation of $E_P$ for 
$\varepsilon_0/\alpha^2 > 100$ has turned out to be difficult, the asymptotic
value of $E_P$ can nevertheless be obtained analytically from the strong SO limit of the
exponential-Bessel expression for $E[\mathbf{\Psi}]$:
\begin{equation}
\label{expBes}
\frac{E[\mathbf{\Psi}]}{\alpha^2\omega_0}=\frac{\tilde{a}^2+\tilde{b}^2}{2}
-\tilde{b}\tilde{\gamma}-\frac{\pi}{\sqrt{2}}\tilde{a},
\end{equation}
whose minimum is at $\tilde{b}=\tilde{\gamma}$ and $\tilde{a}=\pi/\sqrt{2}$,
leading to
\begin{equation}
\label{expBesinf}
\lim_{\varepsilon_0/\alpha^2\rightarrow \infty}\frac{E_P}{\alpha^2\omega_0}
=-\frac{\pi^2}{4}\simeq -2.467.
\end{equation}

\textit{Gaussian ansatz}. The results obtained by using a Gaussian wave function
of the form $f(r)=\mathcal{A}\exp(-a^2r^2)$ are plotted in Fig. \ref{fig3}(b).
Compared to the exponential wave function,
the Gaussian ansatz gives an overall lowering of the polaron energy
for both sinuisodal and Bessel forms of the spinors. 
In the $\varepsilon_0/\alpha^2\ll 1$ limit, and independently of which particular
spinor is used, the ground state polaron energy is found to be:
\begin{equation}
\label{gau1}
\frac{E_P}{\alpha^2\omega_0}=
-\frac{\pi}{8}-\frac{\varepsilon_0}{\alpha^2}
+\mathcal{O}\!\left(\frac{\varepsilon_0^2}{\alpha^4}\right),
\end{equation}
confirming in this regime the linear dependence on the SO coupling of Eq.\eqref{expo2}.
For larger values of the SO coupling, and contrary to the case shown 
in Fig. \ref{fig3}(a), the sinuisodal and Bessel-type
spinors give basically the same values of $E_P$ for all SO couplings up to 
$\varepsilon_0/\alpha^2\simeq 1$. Beyond this value, as for the case with 
the exponential wave function, the polaron energy obtained
from the sinuisodal ansatz becomes larger than that obtained from the Bessel
spinor and, as shown in the inset of Fig. \ref{fig3}(b), rapidly increases while
the Gaussian-Bessel anstaz gives a monotonous lowering of $E_P$.
For $\varepsilon_0/\alpha^2\gg 1$, the
Gaussian-Bessel energy functional has the same form of Eq.\eqref{expBes} with the
latter term substituted by $-2.279\tilde{a}$, which implies 
\begin{equation}
\label{gauBesinf}
\lim_{\varepsilon_0/\alpha^2\rightarrow \infty}\frac{E_P}{\alpha^2\omega_0}
\simeq -2.579.
\end{equation}

\textit{Pekar-type ansatz}.
Let us now evaluate $E_P$ by using in Eqs.\eqref{ansatz} and \eqref{ansatzB}
the Pekar-type ansatz $f(r)=\mathcal{A}(1+a_1 r+a_2 r^2)\exp(-ar)$. 
For zero SO coupling, this ansatz gives $E_P/\alpha^2\omega_0\simeq-0.4046$,\cite{devre1}
which is a lower energy than those obtained from the exponential and Gaussian trial wave functions
and only $0.03\%$ higher than the exact result $-0.40474$ of Ref.[\onlinecite{qinghu}].
As shown in Fig. \ref{fig3}(c), the Pekar-type ansatz gives slightly better estimates of $E_P$
also for nonzero SO couplings, with an overall behavior similar to the previous cases.
Namely, in the weak SO regime one finds
\begin{equation}
\label{pekar1}
\frac{E_P}{\alpha^2\omega_0}=
-0.4046-\frac{\varepsilon_0}{\alpha^2}
+\mathcal{O}\!\left(\frac{\varepsilon_0^2}{\alpha^4}\right),
\end{equation}
and, as before, for stronger SO couplings the energy obtained from the sinuisodal spinor
increases indefinitely with $\varepsilon_0/\alpha^2$.
However, contrary to the exponential and Gaussian ansatzes, the Pekar-type wave function
may give a lower polaron energy when used in combination with the sinuisodal spinor.
This holds true as long as $\varepsilon_0/\alpha^2 \lesssim 2.72$, while for stronger
SO couplings it is the Bessel-type spinor which gives the lower $E_P$ 
[inset of Fig. \ref{fig3}(c)]. A numerical minimization of the asymptotic limit of 
the Pekar-Bessel functional for  $\varepsilon_0/\alpha^2\gg 1$ gives
\begin{equation}
\label{PekBesinf}
\lim_{\varepsilon_0/\alpha^2\rightarrow \infty}\frac{E_P}{\alpha^2\omega_0}
\simeq -2.91,
\end{equation}
which is lower than the asymptotic values of Eqs.\eqref{expBesinf} and \eqref{gauBesinf}.

The results plotted in Fig. \ref{fig3} clearly demonstrate that, since the variational
method provides an upper bound for true ground state polaron energy, the lowering of $E_P$
induced by the SO coupling is a robust feature of the strong coupling Fr\"ohlich-Rashba polaron.
Among the different ansatzes studied, the lower polaron energy is obtained by
using a Pekar-type wave function for $f(r)$ in combination with the sinuisodal spinor for
weak to moderate values of $\varepsilon_0/\alpha^2$ or with the Bessel-type spinor for 
stronger SO couplings. Given that, as discussed above, reasonable values of 
$\varepsilon_0/\alpha^2$ for strongly-coupled polarons fall in the range 
$0 \leq \varepsilon_0/\alpha^2 \lesssim 1-2$, the Pekar-sinuisodal wave function
provides therefore the best description of the Fr\"ohlich-Rashba polaron in this regime.

\subsection{effective mass}
\label{masseff}

\begin{figure*}[t]
\protect
\includegraphics[scale=0.85,clip=true]{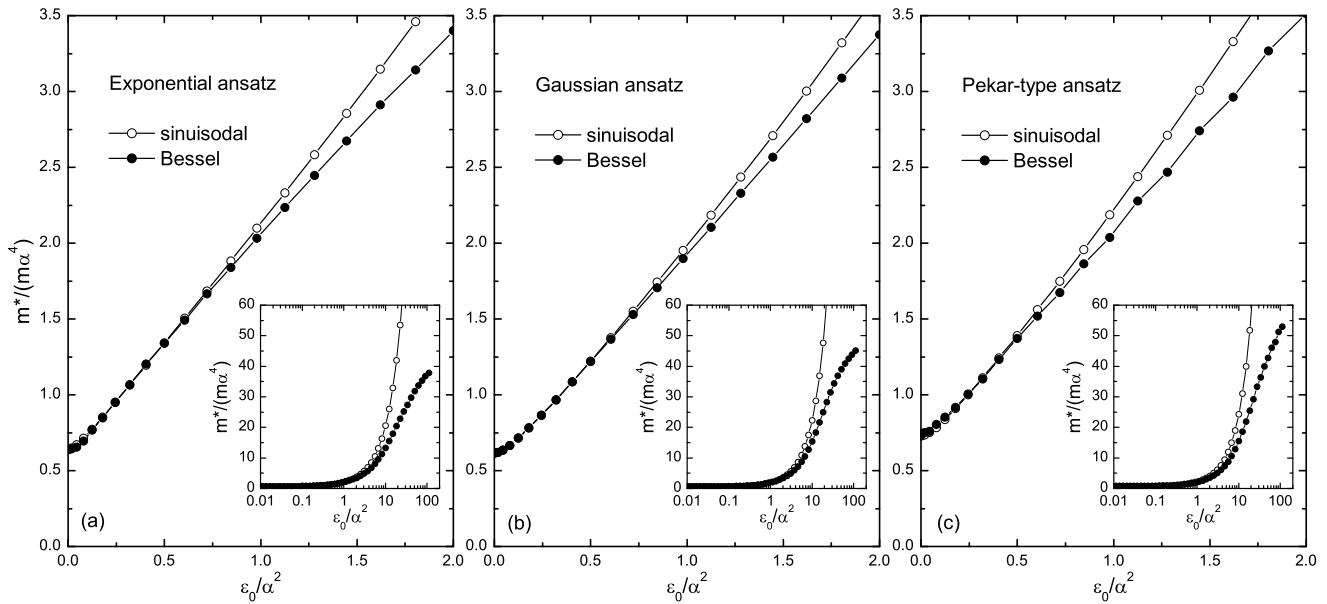}
\caption{Polaron mass enhancement $m^*/m$ in units of $\alpha^4$ as a function of $\varepsilon_0/\alpha^2$
for different ansatz wave functions. (a): exponential; (b): Gaussian; (c): Pekar.
Inset: $m^*/m\alpha^4$ is plotted for a wider range of SO values.} \label{fig4}
\end{figure*}

As demonstrated in Sec.\ref{weak}, the effective mass $m^*$ of a weakly-coupled
polaron is enhanced by the SO interaction and, given the results above,
the same phenomenon is reasonably expected to occur also for the strong-coupling case.
To quantify the polaron mass enhancement within the localized wave function formalism,
it is useful to follow the approach of Refs.[\onlinecite{allcock,parker,providencia}],
briefly described below, where a moving wave packet is constructed from the localized
wave function. The quantity to minimize is
\begin{equation}
\label{HP}
J_{\boldsymbol{\upsilon}}[\mathbf{\Psi}',\xi']=\langle\mathbf{\Psi}',\xi'\vert H
-\boldsymbol{\upsilon}\cdot {\bf P}\vert\mathbf{\Psi}',\xi'\rangle,
\end{equation}
where $\boldsymbol{\upsilon}$ is a Lagrange multiplier, which will turn out to be the mean polaron
velocity, and ${\bf P}={\bf p}+\sum_{\bf q}{\bf q}\,a^\dagger_{\bf q}a_{\bf q}$ is the total
momentum operator. The wave function $\vert\mathbf{\Psi}',\xi'\rangle$ is given by the
product $\mathbf{\Psi}'({\bf r})\vert \xi'\rangle$ where
\begin{equation}
\label{psi1}
\mathbf{\Psi}'({\bf r})=e^{i{\bf p}_0\cdot{\bf r}}\mathbf{\Psi}({\bf r})
\end{equation}
is the electron wave packet with ${\bf p}_0$ being a variational momentum,
$\mathbf{\Psi}({\bf r})$ is the ansatz localized wave function, and
$\vert \xi'\rangle=\mathcal{N}e^{\sum_{\bf q}\xi_{\bf q}'a^\dagger_{\bf q}}\vert 0\rangle$.
Minimization of \eqref{HP} with respect to $\xi_{\bf q}'$ gives now the functional
\begin{align}
\label{HP2}
J_{\boldsymbol{\upsilon}}[\mathbf{\Psi}']
=&\langle\mathbf{\Psi}'\vert H_{el}-\boldsymbol{\upsilon}\cdot{\bf p} \vert\mathbf{\Psi}'\rangle
\nonumber \\
&-\vert M_0\vert^2\int\!\frac{d{\bf q}}{(2\pi)^2}\frac{\vert\rho({\bf q})'\vert^2}{q}
\frac{1}{\omega_0-{\bf q}\cdot\langle\mathbf{\Psi}'\vert\boldsymbol{\upsilon}\vert\mathbf{\Psi}'\rangle},
\end{align}
where ${\bf p}$ is the electron momentum operator and
$\rho({\bf q})'=\langle\mathbf{\Psi}'\vert e^{i{\bf q}\cdot{\bf r}}\vert\mathbf{\Psi}'\rangle$.
By using Eq.\eqref{psi1}, it is easily shown that $J_{\boldsymbol{\upsilon}}[\mathbf{\Psi}']$ reduces to
\begin{align}
\label{HP3}
J_{\boldsymbol{\upsilon}}[\mathbf{\Psi}']=&\langle\mathbf{\Psi}\vert H_{el}\vert\mathbf{\Psi}\rangle
+\frac{p_0^2}{2m}-{\bf p}_0\cdot\boldsymbol{\upsilon} \nonumber \\
&-\vert M_0\vert^2\int\!\frac{d{\bf q}}{(2\pi)^2}\frac{\vert\rho({\bf q})\vert^2}{q}
\frac{1}{\omega_0-{\bf q}\cdot\boldsymbol{\upsilon}},
\end{align}
where $\rho({\bf q})=\langle\mathbf{\Psi}\vert e^{i{\bf q}\cdot{\bf r}}\vert\mathbf{\Psi}\rangle$.
Equation \eqref{HP3} is minimized with respect to ${\bf p}_0$ by setting
${\bf p}_0=m\boldsymbol{\upsilon}$ and, by expanding
the last term of Eq.\eqref{HP3} up to the second order in $\boldsymbol{\upsilon}$, the corresponding
minimum $J_\upsilon[\mathbf{\Psi}]$ becomes:\cite{allcock}
\begin{equation}
\label{HP4}
J_\upsilon[\mathbf{\Psi}]=E[\mathbf{\Psi}]
-\frac{m}{2}\upsilon^2\!\left[1\!+\!\frac{2\vert M_0\vert^2}{m\omega_0^3}
\!\int\!\!\frac{d{\bf q}}{(2\pi)^2}\frac{({\bf q}\cdot\hat{{\bf u}})^2}{q}\vert\rho({\bf q})\vert^2\right],
\end{equation}
where $E[\mathbf{\Psi}]$ is given in Eq.\eqref{sc2}. From the above expression, it
is clear that $J_\upsilon[\mathbf{\Psi}]$ differs from $J_0[\mathbf{\Psi}]$ at least
to order $\upsilon^2$. Hence, if $\mathbf{\Psi}_\upsilon$ and $\mathbf{\Psi}_0$ are the wave
functions which minimize $J_\upsilon[\mathbf{\Psi}]$ and $J_0[\mathbf{\Psi}]$, respectively,
then the difference $\mathbf{\Psi}_\upsilon-\mathbf{\Psi}_0$ is also of order $\upsilon^2$.
As a consequence, the minimum of \eqref{HP4}, $J_\upsilon[\mathbf{\Psi}_\upsilon]$, differs
from $J_\upsilon[\mathbf{\Psi}_0]$ only to order
$(\mathbf{\Psi}_\upsilon-\mathbf{\Psi}_0)^2=\mathcal{O}(\upsilon^4)$ so that,
by neglecting terms of higher order than $\upsilon^2$, minimization of
\eqref{HP4} is achieved by the best wave function which minimizes $E[\mathbf{\Psi}]$.
Therefore, by using $E[\mathbf{\Psi}_0]=E_P-E_0$ and evaluating
$\langle\mathbf{\Psi}_0\vert {\bf P}\vert\mathbf{\Psi}_0\rangle$, from Eqs.\eqref{HP}
and \eqref{HP4} it turns out that
\begin{equation}
\label{EV}
E_P(\upsilon)=E_P+\frac{m}{2}\upsilon^2\!\left[1\!+\!\frac{2\vert M_0\vert^2}{m\omega_0^3}
\!\int\!\!\frac{d{\bf q}}{(2\pi)^2}\frac{({\bf q}\cdot\hat{{\bf u}})^2}{q}\vert\rho_0({\bf q})\vert^2\right],
\end{equation}
permitting us to identify the quantity within square brackets as the mass enhancement
factor $m^*/m$. By integrating over the direction of ${\bf q}$ and by using \eqref{Mq},
$m^*/m$ becomes in the strong-coupling limit
\begin{equation}
\label{massS}
\frac{m^*}{m}=\frac{\sqrt{2}\pi\alpha}{(m\omega_0)^{3/2}}\int_0^\infty\!\frac{dq}{2\pi}q^2
\vert\langle\mathbf{\Psi}_0\vert e^{i{\bf q}\cdot{\bf r}}\vert\mathbf{\Psi}_0\rangle\vert^2,
\end{equation}
which, by replacing the momentum variable by the dimensionless quantity $\tilde{q}=q\ell_P$,
gives a mass enhancement proportional to $\alpha^4$ in the zero SO case. 
By using the exponential, Gaussian,
and Pekar-type ansatzes in Eq.\eqref{massS}, the resulting mass enhancement factor becomes
$m^*/m=(3/16)^3\pi^4\alpha^4\simeq 0.6421\alpha^4$, $m^*/m=(\pi/4)^2\alpha^4\simeq 0.617\alpha^4$,
and $m^*/m\simeq 0.73\alpha^4$, respectively.\cite{notemass}

The results for nonzero SO coupling are plotted in Fig.\ref{fig4} for the sinuisodal (open
circles) and Bessel (filled circles) spinors evaluated with exponential (a), Gaussian (b),
and Pekar-type (c) wave functions. For all cases, $m^*/m$ increases with $\varepsilon_0/\alpha^2$
without much quantitative differences between the various ansatzes as long 
as $\varepsilon_0/\alpha^2\lesssim 2$.
As shown in the insets of Fig. \ref{fig4}, for larger values of the SO coupling the use of the 
sinuisodal spinor largely overestimates the increase of the effective mass compared to the
Bessel-type spinor results. However, despite of the weaker enhancement of $m*/m$, the Bessel-type 
spinors give nevertheless an infinite effective mass at $\varepsilon_0/\alpha^2= \infty$.
Indeed, independently of the particular form of $f(r)$, for 
$\varepsilon_0/\alpha^2\rightarrow \infty$ the expectation value 
$\langle\mathbf{\Psi}_0\vert e^{i{\bf q}\cdot{\bf r}}\vert\mathbf{\Psi}_0\rangle$ 
appearing in Eq.\eqref{massS} goes like $a/q$ for $q\rightarrow\infty$, rendering the
integral over $q$ of Eq.\eqref{massS} divergent.

\section{discussion and conclusions}
\label{concl}

\begin{figure}[t]
\protect
\includegraphics[scale=0.75,clip=true]{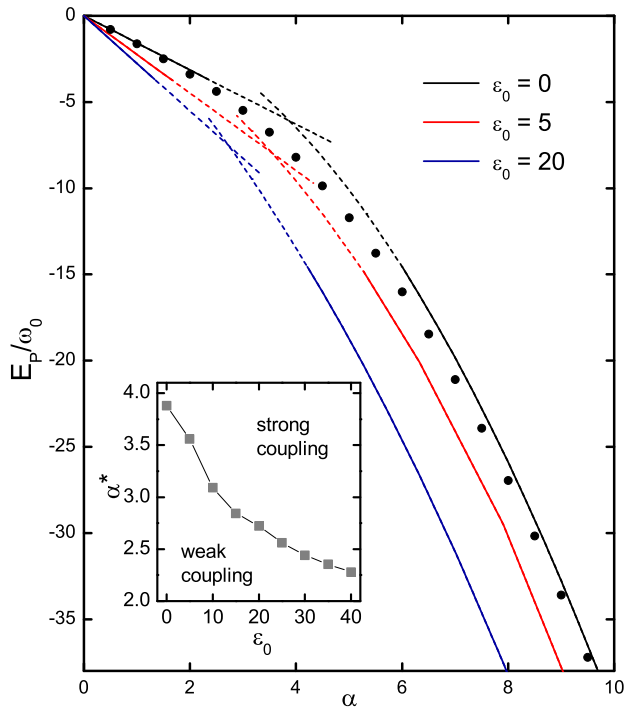}
\caption{(Color online). Ground state polaron energy $E_P$ as a function of the
el-ph coupling $\alpha$ for different values of the dimensionless SO parameter
$\varepsilon_0=E_0/\omega_0$. The straight lines at small $\alpha$ refer to the
weak coupling results, while the curves at large $\alpha$ are the solution of the
strong-coupling theory. The filled circles are the solution of the Feynman path integral
ansatz (see text). The point of intersection between the weak and strong coupling curves
is a measure of the crossover el-ph coupling $\alpha^*$. Inset: $\alpha^*$
is plotted as a function of $\varepsilon_0$.  } \label{fig5}
\end{figure}

The results presented in the previous sections consistently show
that, for both the weak and strong coupling limits of the el-ph
interaction, the ground state energy $E_P$ of the
Fr\"ohlich-Rashba polaron is lowered by the SO interaction and
the mass is enhanced, leading to the conclusion that the Rashba
coupling amplifies the polaronic character. This scenario suggests
also that a weak-coupling polaron at $\varepsilon_0=0$ may be
turned into a strong-coupling one for $\varepsilon_0 > 0$ or, more
generally, that the crossover between weakly and strongly coupled
polarons may be shifted by the SO interaction. This possibility
can be tested by looking at the curves plotted in the main panel
of Fig. \ref{fig5}, where the weak and strong coupling results
for $E_P/\omega_0$ are reported as a function of the el-ph
coupling $\alpha$ for different $\varepsilon_0$ values. For
$\varepsilon_0=0$, the polaron energy follows $E_P/\omega_0\simeq
-\pi\alpha/2$ for small $\alpha$ and $E_P/\omega_0\simeq
-0.4046\alpha^2$ for large $\alpha$. These two limiting behaviors
are plotted in Fig. \ref{fig5} by the uppermost curves and
compared with a numerical solutions of the Feynman variational
path integral for the 2D polaron (filled circles). The largest
deviation of the path integral solutions from the weak and strong
coupling approximations falls in the range of intermediate values
of $\alpha$ and signals a region of crossover between the weakly
and strongly coupled polaron. A rough estimate of the crossover
position is given by a ``critical" coupling, say $\alpha^*$,
obtained by equating the weak and strong coupling results. For
$\varepsilon_0=0$ therefore one has $\pi\alpha/2=0.4046\alpha^2$,
which gives $\alpha^*\simeq 3.9$. Now, as shown in Fig.
\ref{fig5} for $\varepsilon_0=5$ and $\varepsilon_0=20$, the
increase of the SO interaction systematically reduces, for fixed
$\alpha$, the polaron ground state energy and, at the same time,
shifts the intersection point between the weak and strong coupling
curves towards smaller values of the el-ph interaction. The ``critical"
value $\alpha^*$ of the crossover is therefore reduced by the SO 
interaction. For $\varepsilon_0=5$ and $\varepsilon_0=20 $ it is found 
that $\alpha^*\simeq 3.6$ and $\alpha^*\simeq 2.7$, respectively.
The systematic reduction of the crossover coupling by the SO interaction
is made evident in the inset of Fig. \ref{fig5}, where $\alpha^*$ is plotted as a
function of $\varepsilon_0$. From Fig. \ref{fig5} it is also expected that, 
beside the reduction of $\alpha^*$, the
crossover region is likely to be narrowed by $\varepsilon_0$. Indeed,
the intersection between the weak and strong coupling solutions for $\varepsilon_0=20$ is
apparently smoother than the case for $\varepsilon_0=0$, suggesting that the true
ground state energy would deviate less, and in a narrower region around $\alpha^*$,
from the weak and strong coupling solutions.

The scenario illustrated above, and in particular the SO effect
on the crossover coupling, may be verified by quantum Monte-Carlo
calculations of the Fr\"ohlich-Rashba action or, more simply, by
generalizing the Feynman ansatz for the retarded
interaction to $\varepsilon_0>0$.\cite{feynman} The results presented here
on the limiting cases $\alpha\ll 1$ and $\alpha\gg 1$ may then serve as 
a reference for such more general calculations schemes for arbitrary values
of the el-ph coupling and of the SO interaction.

Let us discuss, before concluding, possible generalizations of the Fr\"ohlich-Rashba
model employed here and the consequences on the polaronic character.
Let us remind that in Ref.[\onlinecite{cappe}] it has been demonstrated that also
for a momentum independent el-ph interaction model, the Rashba SO term leads
to an effective enhancement of the el-ph coupling. The SO induced lowering of the 
polaron ground state is therefore robust against the specific form of the 
el-ph interaction, so that a similar behavior is expected to occur also
when considering the contributions from interface or surface phonon modes.
However, a different form of the SO interaction term may lead to a much weaker effect.
Consider for example the situation in which, in addition to
the Rashba SO coupling, the system lacks also of bulk inversion symmetry, as in III-V
semiconductor heterostructures, leading to an extra SO term of the Dresselhaus 
type.\cite{zutic,dressel} When both SO contributions are present, the square root
divergence of the DOS at the bottom of the band of the free electron disappears, and
it is replaced by a weaker logarithmic divergence at higher energies. In this situation
therefore, at least for weak el-ph couplings, the SO interaction is expected to have a
weaker effect on the polaron ground state, which tends to vanish as the Dresselhaus
term becomes comparable to the Rashba one. 

Let us conclude by noticing that, recently, 
the possibility of varying the coupling of 2D Fr\"ohlich polarons 
in a controlled way has been experimentally demonstrated by acting on the dielectric 
polarizability of organic field-effect transistors.\cite{fratini} The results  
presented here suggest that tunable 2D Fr\"ohlich polarons may be achieved
also by acting on the SO coupling, which can be tuned by applied gate
voltages in quasi-2D structured materials.

\acknowledgements
The author thanks Emmanuele Cappelluti and Frank Marsiglio for valuable comments.


\begin{thebibliography}{99}
\bibitem{noteself}
The form of the self-energy reported in Eq.\eqref{self2} is not limited to
the weak-coupling limit. For a more general derivation 
see Ref.[\onlinecite{cappe}].
\bibitem{massholstein}
There are however quantitative differences. Indeed,
in the strong SO limit, the Holstein-Rashba model gives an effective
mass enhancement proportional to $\sqrt{\varepsilon_0}$, {i.e.}, a weaker
increase with respect to the linear dependence found here for the Fr\"ohlich-Rashba
model (see Ref.[\onlinecite{cappe}]).
\bibitem{note_trap}
The term ``trapping" is used here in a loose way, since no
real trapping effect exist for the continuum Fr\"ohlich polaron.
\bibitem{notemass}
Note that the result for the exponential ansatz differs from that of
Ref.[\onlinecite{devre2}]. The difference stems from an incorrect
expression of the exponential decay factor $a$: in Ref.[\onlinecite{devre2}]
$a/\ell_P=3\pi/8$ while instead it should be given by $a/\ell_P=3\pi/16$, as in
Ref.[\onlinecite{devre1}] and in the present paper.
\end{thebibliography}
\end{document}